# Saturated absorption at nanowatt power levels using metastable xenon in a high-finesse optical cavity

G. T. Hickman,[*] T. B. Pittman, and J. D. Franson

*Department of Physics, University of Maryland Baltimore County, Baltimore, MD 21250, USA*
[*]hickman1@umbc.edu

**Abstract:** Strong saturated absorption at nanowatt power levels has been demonstrated using metastable xenon in a high-finesse optical cavity. The use of metastable xenon allows a high quality factor of $Q=2\times10^8$ to be achieved at relatively high atomic densities without any contamination or damage to the optical surfaces, which is often a problem when using high-density rubidium or other alkali atoms. This technique provides a relatively straightforward way to produce nonlinearities at the single-photon level with possible applications in quantum communications and computing.

## 1. Introduction

Optical nonlinearities at ultra-low power levels have many potential applications in quantum optics, including quantum computing and communications [1–3]. Nonlinearities of this kind can be produced in a number of ways [4–7], such as by trapping a single atom in a micro-cavity with a small mode volume [8-10]. Here we report the results of an experiment in which strong saturated absorption was demonstrated at nanowatt power levels using a gas of metastable xenon in a high-finesse cavity. The use of metastable xenon avoids the contamination of the optical surfaces that can occur when using high-density alkali vapors [11].

The strong nonlinearity demonstrated in this way is consistent with the ability to generate other forms of optical nonlinearities, such as single-photon cross-phase modulation [12], which would have a number of practical applications [1–3]. It was previously shown that comparable cross-phase modulation can be obtained using either a gas of atoms or a single trapped atom [13]. The technique described here provides a relatively simple method for producing optical nonlinearities at the two-photon level without the need to trap single atoms.

The remainder of the paper begins with a description of the relevant properties of metastable xenon in Section 2. Section 3 describes the resonant cavity and xenon source used in the experiments. Section 4 outlines the experimental setup for the saturated absorption measurements and presents the results. A summary and conclusion are presented in Section 5.

## 2. Metastable Xenon

Metastable xenon has many properties similar to those of Rb, such as level structure and oscillator strength. In this section we provide a brief description of some of the properties important for experiments in nonlinear optics.

An overview of the relevant xenon energy levels and transitions of interest is shown in Figure 1. The metastable $6s[3/2]_2$ state of xenon can be used as an effective ground state for nonlinear optical experiments due to its long intrinsic lifetime of ~ 43 s [14]. Once excited to the metastable state, a pair of transitions in a ladder-type configuration becomes available: a transition from $6s[3/2]_2$ to $6p[3/2]_2$ at 823 nm and another from $6p[3/2]_2$ to $8s[3/2]_2$ at 853 nm. These have dipole matrix elements comparable to those of transitions commonly used in Rb. Only the first transition is used for the saturated absorption measurements reported here, while the second transition could be used to produce nonlinear phase shifts in subsequent experiments.

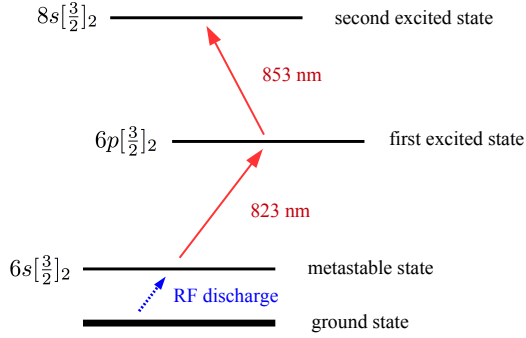

FIG. 1. Xenon energy level diagram showing the energy levels and transitions of interest. An RF discharge populates the long-lived $6s[3/2]_2$ metastable state which is then used as an effective ground state. A pair of transitions at 823 nm and 853 nm is then available for nonlinear optics experiments.

Nanowatt saturation of the $6s[3/2]_2$ to $6p[3/2]_2$ transition has been previously observed using tapered optical fibers [15]. However, the use of a cavity may be more advantageous for producing the single-photon nonlinear phase shifts for use in quantum information processing [12, 16].

Spectroscopy of the $6s[3/2]_2$ to $6p[3/2]_2$ transition in xenon yields multiple absorption dips, due to the presence of various isotopes and hyperfine splittings as illustrated in Figure 2. Our saturated absorption measurements involve four of the largest dips. Three of these are caused by hyperfine components in xenon 129 and 131. The remaining dip is produced by the combined absorption from all even xenon isotopes, which cannot be individually resolved because of Doppler broadening. The measured absorption dips and their associated hyperfine levels will be shown in Section 4.

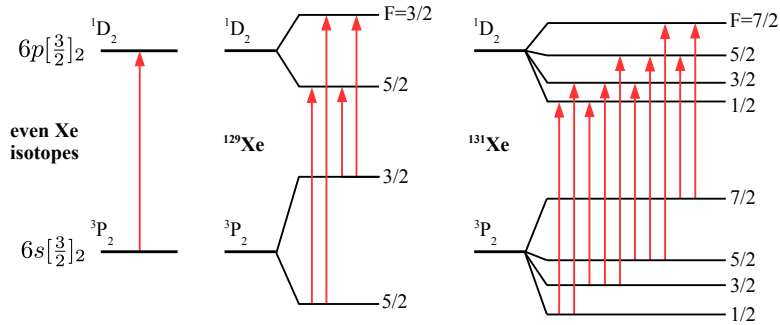

FIG. 2. Hyperfine splittings for the $6s[3/2]_2$ to $6p[3/2]_2$ transition in the naturally occurring xenon isotopes, as shown in reference [17]. Each of the 7 even isotopes has a nuclear spin of 0, leading to a total of 21 hyperfine components.

### 3. Resonant Cavity

Figure 3 illustrates the cavity and metastable xenon source used in this experiment. The cavity consisted of a pair of super-polished dielectric mirrors with 2.5 cm radii of curvature in a confocal configuration, with a mode-field beam diameter of 116 μm at the cavity center. The measured finesse was approximately 4,000, the observed quality factor was $Q=2\times10^8$, and the measured linewidth at 823 nm was approximately 1.5 MHz. A moderate value of the finesse was chosen so that the cavity loss would be dominated by transmission through the mirrors rather than by scattering or other loss mechanisms. A pair of rectangular "wings" with insets to house the mirrors was machined into a solid cylindrical nickel block. A pair of small holes drilled through the wings provided space for a beam to couple into and out of the cavity's fundamental transverse mode.

The resonant frequency of the cavity was tuned by controlling the temperature of the nickel block. It was fastened at the bottom onto a copper feedthrough rod which supported the block and provided thermal contact to a heater located outside of the vacuum chamber (a 4.5-inch stainless steel cube). The resonant frequency of the cavity could be shifted by a full free spectral range (6 GHz) with a change in temperature of only a fraction of a degree C. We avoided the use of piezoelectric elements for cavity tuning due to the presence of the RF discharge described below and for improved stability.

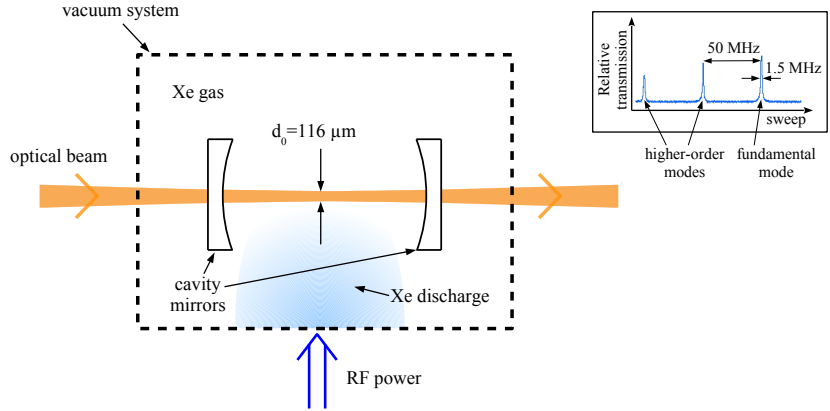

FIG. 3. Conceptual overview of the resonant cavity and xenon source. Xenon was excited into the metastable state via an RF discharge. The fundamental transverse cavity mode could be easily isolated from the higher-order modes, as shown in the inset in an oscilloscope trace of the cavity transmission spectrum (for this trace, the cavity was deliberately misaligned to highlight the higher-order modes).

The resonant cavity was cleaned using standard ultra-high vacuum (UHV) techniques and placed into the vacuum chamber. The chamber was then filled with approximately 0.1 torr of naturally-occurring xenon and 0.9 torr of helium buffer gas after pumping out the air with a turbopump to a pressure of approximately $10^{-8}$ torr.

The xenon metastable state can be populated using an RF or DC discharge, which can produce metastable densities as high as $10^{13}$ cm$^{-3}$ [18]. The density of metastable xenon used in this experiment was roughly $10^9$ cm$^{-3}$. A resonant RF circuit consisting of a coil and capacitors was placed against a glass window flange on the outside of the chamber. RF power at a frequency of 130 MHz was applied to produce an electrical discharge. Metastable xenon created in the discharge region diffused through the vacuum chamber into the region between the two mirrors. The mirrors were exposed to the xenon discharge through the 2 mm diameter holes used to couple the beam, but they showed no measurable degradation or reduced finesse as a result of the discharge.

### 4. Saturated Absorption Measurements

Figure 4 provides an overview of the optics and control electronics used in the experiment. The output of a frequency-stabilized diode laser (Toptica DL 100 pro design) tuned to 823 nm was coupled into an optical fiber and split using a fiber coupler. The additional signal was fed into a precision wavelength meter that continually monitored the laser frequency. Light polarization in the fiber was not controlled since the xenon absorption was not polarization sensitive. The beam was further divided into a high-power and a low-power beam, with the intensity in each arm controlled by in-line fiber optic attenuators. The high-intensity beam was required to saturate the absorption while locking the laser frequency to the cavity transmission peak, since there was very little transmission through the cavity at low intensities when on resonance with the atomic transition. The low intensity beam was then used to perform the actual experiment.

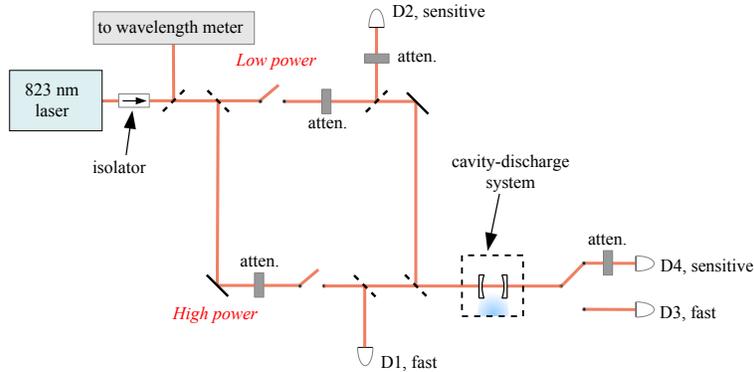

FIG. 4. Overview of the fiber-based apparatus used to measure saturated absorption in the $6s[3/2]_2$ to $6p[3/2]_2$ transition of metastable xenon. Variable attenuators and fiber-coupled switches allowed the system to switch quickly between high-intensity for locking the laser to the cavity resonance and low-intensity for probing the xenon absorption.

The input into each beam was switched on or off using a pair of fiber-coupled switches (Thorlabs model OSW12-780E) with a switching time of ~ 1 ms, making it possible to quickly change between a high-power or low-power cavity input beam. Another fiber-coupled switch was used at the cavity output to choose between a fast photoreceiver (D3) to measure the high-intensity transmission or a more sensitive detector (D4) to measure the low-intensity signal. Two more detectors (D1 and D2) monitored the high-power and low-power beam intensities in order to compensate for effects due to laser intensity fluctuations.

Once the peak of the fundamental transverse cavity mode was coarsely located, the laser frequency was swept across it in order to map out the shape of the transmission curve, which was used to find the point of maximum transmission. This scan was performed first with the high-intensity beam and then the low-intensity probe beam. Because the high-intensity beam fully saturated the absorption, the ratio of transmission values taken at the two power levels indicated the optical depth of the intra-cavity medium at the lower intensity. A Labview routine controlled the laser and the switch states to maintain the frequency lock on cavity resonance and to continuously measure the xenon absorption at the low power level.

The temperature of the cavity was varied in such a way that the cavity resonance was slowly scanned across the metastable xenon absorption spectrum at 823 nm. This process was repeated multiple times using different intensities of the weaker beam in order to map out the absorption profile as a function of frequency for each intensity value.

Figure 5 shows some of the experimental results obtained using low-intensity beams with powers of 0.5, 2 and 19 nW. We found that the absorption depended linearly on the input power at powers below about 0.5 nW. As shown in the figure however, there was a significant decrease in the relative absorption as the power was increased above 0.5 nW. For example, the transmission at 364.097 THz (corresponding to the F=5/2 to F=5/2 hyperfine transition in the 129Xe isotope [19]) increased from 10% to 50% as the power was increased from 0.5 nW to 19 nW.

Because of its very large optical density, the depth of the broad central dip from the even Xe isotopes did not appear to decrease with greater input power. In this case the decrease in relative absorption manifested itself in an effective narrowing of the apparent width of the dip as input power was increased. When the absorption spectrum was measured with an input power of tens of µW from the high-power beam, transmission was flat with no noticeable absorption.

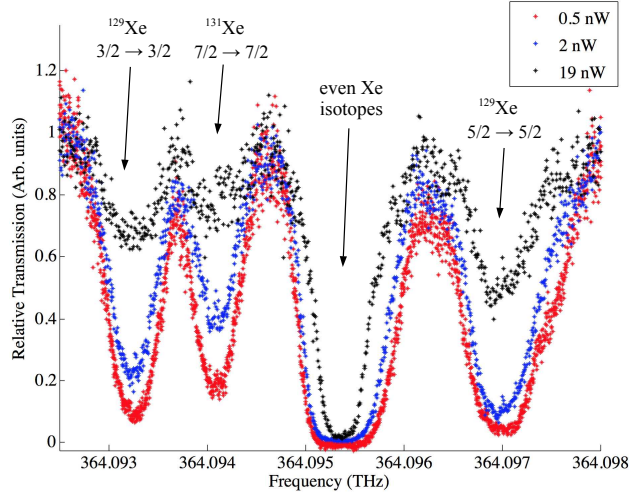

FIG. 5. Absorption data for intra-cavity metastable xenon with input powers of 0.5, 2 and 19 nW. The hyperfine components that dominate the spectrum are labeled by isotope and $F_i \rightarrow F_f$ [19, 20]. A significant change in the absorption can be seen between power levels of 0.5 and 2 nW.

Hyperfine pumping can lead to apparent differences in saturation characteristics for the various isotopes and hyperfine components [21]. However, the single dip caused by the even xenon isotopes is not affected by hyperfine pumping since each of these isotopes has a nuclear spin of 0, though pumping of the $6s[3/2]_1$ state remains a possibility. For each of the visible dips the highly nonlinear power-dependent absorption indicates a saturation effect at nW power levels. Saturated absorption at these low power levels demonstrates the large nonlinearities possible with this system.

### 5. Summary and Conclusions

In conclusion, we have experimentally demonstrated saturated absorption at nanowatt power levels using metastable xenon inside a high-finesse cavity. The use of metastable xenon avoids the potential difficulties associated with the use of high-density rubidium or other alkali vapors, which often degrade the optical surfaces over time [11]. In contrast, the metastable xenon did not produce any observable degradation in the optical surfaces or the cavity finesse.

The observed saturated absorption at these low power levels suggests that this approach should be able to produce other nonlinear optical effects at ultra-low power levels as well. For example, density-matrix calculations show that we should be able to achieve a single-photon cross-phase shift of approximately 20 mrad using our system [12]. That would be sufficient for a number of proposed applications, including quantum computing [2] and quantum key distribution [3]. The relative simplicity and robustness of this approach makes it well-suited for applications of this kind.

The authors would like to acknowledge valuable discussions with C. J. Broadbent, John C. Howell, and A. V. Sergienko. This work was supported by DARPA DSO Grant No. W31P4Q-12-1-0015.